# Ionization waves in low-current DC discharges in noble gases obtained with a hybrid kinetic-fluid model

Vladimir I. Kolobov[1,2] and Robert R. Arslanbekov[2]

[1]The University of Alabama in Huntsville, Huntsville, AL, USA
[2]CFD Research Corporation, Huntsville, AL, USA

**Abstract**

A hybrid kinetic-fluid model is used to study ionization waves (striations) in a low-current plasma column of DC discharges in noble gases. Coupled solutions of a kinetic equation for electrons, a drift-diffusion equation for ions, and a Poisson equation for the electric field are obtained to clarify the nature of plasma stratification in the positive column. A simplified two-level excitation-ionization model is used for the conditions when the nonlinear effects due to stepwise ionization, gas heating, and Coulomb interactions among electrons are negligible. It is confirmed that the nonlocal effects are responsible for the formation of moving striations in DC discharges at low plasma densities and low values of *pR* (the product of gas pressure and tube radius). The calculated properties of self-excited waves of *S, P,* and *R* types in Neon and *S* type in Argon agree with available experimental data. The reason for Helium plasma stability to stratification is clarified. It is shown that sustaining stratified plasma is more efficient than striation-free plasma when the ionization rate is a nonlinear function of the electric field. However, the nonlinear dependence of the ionization rate on the electric field is not required for plasma stratification. Striations of *S*, *P* and *R* types in Neon exist with minimal or no ionization enhancement. Effects of the column length and plasma density on the wave properties are demonstrated.

## I. Introduction

Weakly-ionized collisional plasma is prone to instabilities and pattern formation. The bright and dark layers along the electric current are called striations. Spatial patterns across the current are associated with constriction, filamentation, and streamer formation. Nonlinear and nonlocal effects are responsible for most of these self-organization processes. The nonlinear effects appear because a small fraction of energetic electrons usually controls the ionization processes in plasma. The number of these electrons depends nonlinearly on the ratio of the electric field to gas density, *E/N*. The ionization rate can also depend on the electron density due to stepwise ionization and Coulomb collisions. Specific nonlocal effects arise due to the peculiarities of electron kinetics in collisional plasma [1, 2]. The electron mean free path and the electron energy relaxation length can differ by two orders of magnitude depending on the gas type and the electron kinetic energy [3]. Identifying and understanding specific mechanisms of plasma self-organization in each case remains challenging.

This paper deals with plasma stratification in noble gases. Striations have been observed in gas discharges long before Langmuir introduced the term plasma. The nature of striations as ionization waves became clear by the end of the 70$^{th}$ [4, 5, 6]. Several types of ionization waves in noble gases have been identified depending on gas pressure and discharge current [7, 8]. Studies of electron kinetics in uniform and spatially modulated electric fields were performed using numerical solutions of the Boltzmann equation and particle-based (PIC) methods [9, 10]. These studies contributed to understanding the origin of plasma stratification and helped interpret the famous Frank Hertz experiments [11]. However, the interplay between the nonlinear and kinetic effects responsible for plasma stratification in different gases and discharge types remained unclear.

The most studied are moving striations in DC discharges in noble gases. Several types of waves have been observed depending on the gas pressure $p$, the discharge current $i$, and the tube radius $R$. Similarity laws [12] allowed producing experimental maps of discharge types in the ($pR$, $i/R$) plane shown in Figure 1 for Neon and Argon. Self-excited moving striations have been observed in the white areas of the map. Curve (1) corresponds to the conditions when the Debye radius equals the tube radius. The dark area at low $pR$ and $i/R$ to the left of the curve (1) corresponds to dark (Townsend) discharge with standing striations near the cathode.

The green area in Figure 1 corresponds to a striation-free plasma controlled by ambipolar diffusion to the wall. In this area, artificial striations have been excited to study the dispersion characteristics of the ionization waves. Volume recombination dominates over surface recombination at $pR$ values above the horizontal section of the line (2). The vertical section of line 2 (including its dashed extension to lower $pR$) corresponds to the appearance of nonlinear dependence of the ionization frequency on electron density. The blue area corresponds to a striation-free, diffuse, or radially constricted arc column. Line (3), called the Pupp boundary, corresponds to the transition from glow to arc discharge. No discharges can be maintained at $pR$ values below the curve (4) because the ionization cross section has a finite value. The two horizontal dashed lines indicate gas pressure $p$ at which the electron mean free path $\lambda$ and the electron energy relaxation length $\lambda_u$ is equal to the tube radius $R$, respectively

Several striation types have been identified in the ($pR$, $i/R$) map. Striations in diffuse and constricted discharges near the Pupp boundary (3) appear due to a nonlinear dependence of the ionization frequency on the electron density. Three types of waves have been observed at lower values of $i/R$ below the dashed line $\lambda_u = R$ in Figure 1. The potential drops over striation length for the S, P and R waves have constant values independent of $i/R$ and $pR$, and characterize the gas type (Novak law). These waves appear due to nonlocal electron kinetics under discharge conditions when the electron energy balance is controlled by inelastic collisions with atoms and Coulomb interactions among electrons are negligible. Ionization waves can co-exist with ion sound waves al low values $pR$ between the line $\lambda_u = R$ and curve (4) [14].

Striations in diffuse discharges at medium pressures near and above the dashed line $\lambda_u = R$ are often irregular. They correspond to the discharge conditions when the energy loss in elastic collisions with atoms controls the electron energy balance. The positive column plasma transitions from diffuse to constricted with an increasing discharge current. The striations gradually become two-dimensional with substantial variations of the radial plasma profile along the striation length.

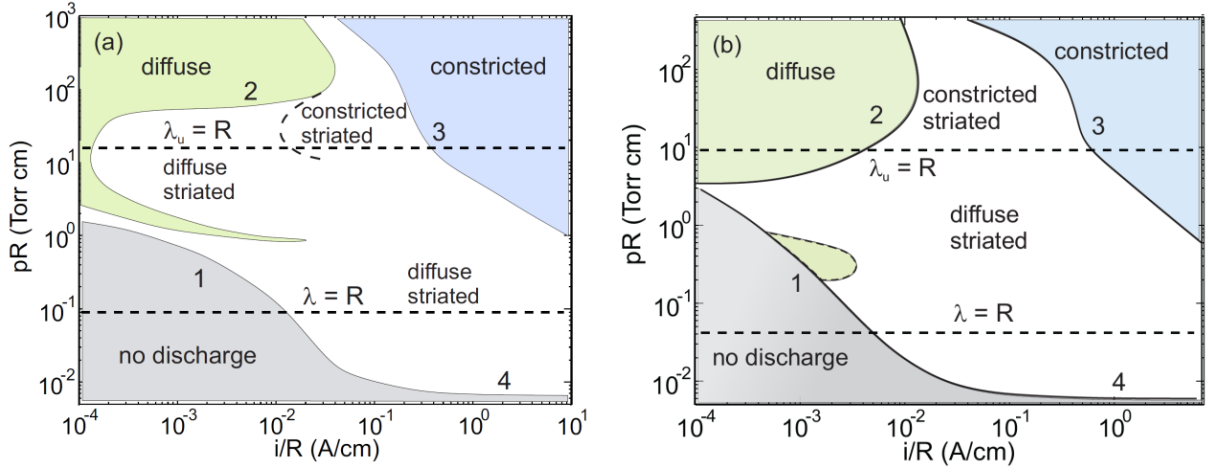

Figure 1: Discharge forms in Neon (a) and Argon (b) for different currents $i$ and gas pressures $p$ (after [12, 13, 14]).

In DC discharges of molecular gases, both moving and standing striations were detected [15]. Standing striations were observed in radio-frequency (RF) discharges [16]. Recently, standing striations have appeared in computer simulations of Capacitively Coupled Plasma (CCP) in $CF_4$ [17] and DC discharges in Nitrogen [18]. Analysis of plasma stratification in molecular gases involves complicated chemistry and the kinetics of vibrationally excited molecules, in addition to electron kinetics.

Recent progress in understanding plasma stratification in noble gases has been achieved using computer simulations with fluid [19, 20], hybrid [21], and Particle-in-Cell (PIC) models [22]. Two-dimensional fluid models have been applied to study ionization waves in DC [19] and RF [20] discharges in Argon at relatively high currents when the nonlinear dependence of the ionization frequency on electron density is responsible for plasma stratification. A hybrid model using a Fokker-Planck kinetic solver for electrons and a fluid model for ions has been developed [21] for standing and moving striations in Argon discharges at low currents when nonlocal effects are the leading cause of stratification. Boeuf [22] used a PIC model to study moving striations in DC discharges in Argon and Neon in the absence of metastable atoms and Coulomb collisions among electrons.

The present paper aims to further clarify the nature of plasma stratification in noble gases at low ionization degrees when the nonlinear dependence of the ionization frequency on plasma density and gas heating are insignificant. We first briefly describe the computational model and boundary conditions. The central part of the paper is devoted to analyzing plasma stratification in a positive column in Argon, Neon, and Helium with periodic boundary conditions. Finally, we briefly discuss the impact of finite column length and near-electrode effects on wave properties and identify a few open questions for future research.

## II. Computational model

The hybrid model used in the present paper was first described in [23] and later adapted in [21] to study striations in Argon DC discharges. Here, we extend this model by implementing periodic boundary conditions similar to [22]. The enhanced model is applied to study the stratification of a long positive column in DC discharges in noble gases.

### 1. Basic equations

The electrons are described via a Fokker-Planck (FP) kinetic equation for the Electron Energy Probability Function (EEPF) $f_0(r,u,t)$ in the $(r,u)$ phase space [24, 25]:

$$\frac{\partial f_0}{\partial t} - \frac{1}{v}\left(\nabla - E\frac{\partial}{\partial u}\right) \cdot vD_r \left(\nabla - E\frac{\partial}{\partial u}\right) f_0 - \frac{1}{v}\frac{\partial}{\partial u}(v\Gamma_u) = C_0 \tag{1}$$

Here, $E$ is the electric field, $v$ is the amplitude of electron velocity, $u = mv^2/(2e)$ is the volt-equivalent of the electron kinetic energy, $D_r = v^2/(3\nu)$ is a diffusion coefficient in the configuration (physical) space, and $\nu(v)$ is the transport collision frequency. The energy flux density, $\Gamma_u$, describes processes associated with small energy changes due to quasi-elastic collisions of electrons with atoms, Coulomb interactions, etc. At $\Gamma_u = 0$, the left-hand side of Eq. (1) describes electron diffusion in the phase space $(r,u)$ over surfaces of constant total energy $\varepsilon = u - \varphi(\mathbf{r})$, where $\varphi(\mathbf{r})$ is the electrostatic potential [21]. The EEPF $f_0$ is normalized on electron density $n_e$ and has units eV$^{-3/2}$ m$^{-3}$.

The right-hand side of Eq. (1) contains inelastic collisions. The excitation of an atomic state with the energy threshold $\varepsilon_1$ is described by

$$C_{exc} = -\nu^*(u)f_0(u) + \frac{\sqrt{u+\varepsilon_1}}{\sqrt{u}}\nu^*(u+\varepsilon_1)f_0(u+\varepsilon_1) \tag{2}$$

where $\nu^*(u)$ is the inelastic collision frequency. The direct ionization by electron impact is described as:

$$C_{ion} = -\nu^{ion}(u)f_0(u) + 4\frac{\sqrt{2u+\varepsilon_{ion}}}{\sqrt{u}}\nu^{ion}(2u+\varepsilon_{ion})f_0(2u+\varepsilon_{ion}) \tag{3}$$

where $\nu^{ion}(u)$ is the ionization frequency, and $\varepsilon_{ion}$ is the ionization threshold. This ionization model assumes that the kinetic energy is evenly distributed between the primary and secondary electrons after an ionization event.

The FP kinetic equation (1) is valid when the Electron Velocity Distribution Function (EVDF) $f_e(\mathbf{r},\mathbf{v},t)$ and the electric field $E$ change slowly on the electron mean free path $\lambda$ during the time $1/\nu$ between collisions [26]. The angular frequency of the ionization waves $\omega$ is always lower than the transport collision frequency $\nu$, so the time constrain is well satisfied. The space constraint is satisfied for the $pR$ values above the horizontal dashed lines $\lambda = R$ in the maps shown in Figure 1. Below the lines $\lambda = R$, the two-term spherical harmonics expansion used for deriving Eq. (1)

becomes invalid. The Boltzmann equation for $f_e(\mathbf{r}, \mathbf{v}, t)$ should be used instead for the analysis of EVDF formation and plasma stratification under these conditions [2].

Below, we assume that plasma is contained in a long cylindrical tube with a length $L > R$. We reduce the kinetic equation (1) for a two-dimensional phase space $(x, u)$:

$$\frac{\partial f_0}{\partial t} - \nabla \cdot (\mathbf{D}_2 \nabla f_0) - \frac{1}{v}\frac{\partial}{\partial u}(v\Gamma_u) = C_0 - C_{wall} , \qquad (4)$$

where the diffusion tensor $\mathbf{D}_2$ is defined as [21]:

$$\mathbf{D}_2 = D_r \begin{pmatrix} 1 & E \\ E & E^2 \end{pmatrix}, \qquad (5)$$

and $E(x)$ denotes the axial electric field.

The radial loss of electrons to the wall in the 1d1u model can be included in the form [27]:

$$C_{wall} = -\frac{1}{3}\left(\frac{v}{R}\right) f_0(u)\Theta(u - \Phi_w) \qquad (6)$$

where $\Phi_w(x)$ is the wall potential with respect to plasma axis, and $\Theta(x)$ is the step function. However, in the present work, we used a simpler expression:

$$C_{wall} = -f_0/\langle \tau_{wall}\rangle = -f_0\langle \nu_i\rangle \qquad (7)$$

where $\langle \nu_i \rangle$ is the space-averaged ionization frequency and $\tau_{wall}$ is the time of electron diffusion to the wall. This model assumes that all electrons are lost radially at the same rate they are created along the striation [22]. Our tests showed that the choice of the radial loss model does not significantly affect the results. More advanced models, such as the one given by Eq. (6), may be used in future studies.

For the numerical solution of the FP kinetic equation, we introduce a computational grid in phase space. The maximal energy $u_{max}$ is selected about 2-3 $\varepsilon_1$. The boundary condition is specified as $f_0(x, u = u_{max}) = 0$. The boundary condition at $u = 0$ ensures the absence of electron flux from this boundary

$$\frac{\partial f_0}{\partial x} - E\frac{\partial f_0}{\partial u} \to 0 \qquad (8)$$

The boundary conditions in space are discussed below.

Ions are described using a drift-diffusion model:

$$\frac{\partial n_i}{\partial t} + \frac{\partial}{\partial x}\left(\mu_i n_i E - D_i \frac{\partial n_i}{\partial x}\right) = I - \frac{n_i}{\tau} \qquad (9)$$

where $\mu_i$ and $D_i$ are the ion mobility and diffusion coefficients, and $I$ is the ionization rate by electron impact. The ion loss term matches the electron loss, $\langle \nu_i \rangle \tau = 1$, where $\langle \nu_i \rangle$ is the ionization frequency averaged over striation length to ensure charge conservation. The electric field $E = -\partial \varphi / \partial x$ is calculated by solving the Poisson equation for the potential $\varphi(x,t)$. We assume that the initial EEPF is Maxwellian and the initial ion density equals electron density.

The coupled set of the FP kinetic equation for electrons (1), the drift-diffusion for ions (9), and the Poisson equation for the electric potential was solved using COMSOL with an implicit (BDF) time-stepping method. A direct (MUMPS) solver was employed at each time step for all quantities. We do not use the logarithmic transformation of the ion drift-diffusion and the FP kinetic equations used in our previous paper [21]. We found that converging the coupled equations with different dimensionality solved simultaneously is substantially better without this transformation (essential for long transient simulations). Small negative values of $f_0$ and $n_i$ do not affect the results.

## 2. Boundary conditions for the positive column model

The positive column is usually considered an autonomous system weakly influenced by near-electrode processes. The electric field in the positive column plasma is established to balance the ionization and the particle loss to the wall due to ambipolar diffusion or volume recombination. The positive column of a finite length can be considered a cavity resonator containing a set of modes representing strata of various types. From this point of view, S-, P-, and R-striations were interpreted as resonances at the fundamental wavelength (S-wave), half-wavelength (P-wave), and 2/3 of the fundamental wavelength (R-wave) [28]. However, the work [28] considered only the electronic component of the plasma. In the present work, we analyze plasma stratification by solving a self-consistent problem, which includes electrons, ions, and a self-consistent electric field.

To study the effects of the column length $L$ on plasma stratification, we apply periodic boundary conditions for the EEPF and the particle fluxes:

$$(f_0)_0 = (f_0)_L \tag{10}$$

$$-(\boldsymbol{D}_2 \nabla f_0 \cdot \hat{n})_0 = (\boldsymbol{D}_2 \nabla f_0 \cdot \hat{n})_L \tag{11}$$

where $\hat{n}$ denotes the unit normal to the boundary. This BC ensures the continuity of the EEPF and the electron flux in phase space.

We also apply periodic boundary conditions for ions by making the ion density and the ion flux equal at the boundaries:

$$(n_i)_0 = (n_i)_L \tag{12}$$

$$-\left(\left(\mu_i n_i E(x) - D_i \frac{\partial n_i}{\partial x}\right)\right)_0 = \left(\left(\mu_i n_i E(x) - D_i \frac{\partial n_i}{\partial x}\right)\right)_L \tag{13}$$

Using periodic BCs implies that the positive column always contains an integer number of waves.

The periodic BCs ensure that the total number of electrons and ions remains the same, provided that electron and ion losses are equal. As the total space charge remains zero (initially, a quasineutral plasma with $n_e = n_i$ is assumed) along the striation length, the periodic BC for the electric field $E(0) = E(L)$ is automatically satisfied.

We prescribe the potential drop $U$ over the column length $L$, which gives the average electric field $\langle E \rangle$ in stratified plasma. In our simulations, the radius of the tube $R$ was computed for a prescribed $\langle E \rangle$ from the balance of ionization and loss:

$$R = 2.4 \sqrt{\frac{D_a}{\langle v_i \rangle}} \tag{14}$$

The average electron temperature $\langle T_e \rangle$ was used to evaluate the ambipolar diffusion coefficient, $D_a$. By gradually changing $U$ we have observed the formation of different types of waves depending on the values of $\langle E \rangle$.

## III. Results of Simulations

Simulations were performed in Argon, Neon, and Helium at pressures 0.1, 0.2, 0.4, and 1 Torr, and plasma densities $\langle n \rangle$ varied from $10^{14}$ to $10^{16}$ m$^{-3}$. The electron cross-section data and ion mobilities used for the three gases were taken from [29]. The potential drop over striation length $\Delta\varphi_\Lambda$ and the average electric field $\langle E \rangle$ were calculated as:

$$\Delta\varphi_\Lambda = \int_0^\Lambda dx E(x,t) = v_s \int_0^T dt E(x,t) = \langle E \rangle \Lambda \tag{15}$$

Here, $v_s$ denotes the propagation velocity of the wave and $T = 2\pi/\omega$ its period. The tube radius $R$ was found self-consistently from Eq. (14) for each value of $\langle E \rangle$.

Figure 2 illustrates the dynamics of the plasma stratification process for the positive column of length $L = 15$ cm in Neon at the pressure of 1 Torr, the average plasma density $\langle n \rangle = 10^{15}$ m$^{-3}$, and the applied voltage $U = 100$ V. The calculated radius of the discharge tube is 5.1 mm. Figure 2a shows the contours of electron density $n_e(x,t)$, and Figure 2b shows the time evolution of the electron density in the middle of the gap, $n_e(L/2,t)$. The dynamics of plasma stratification in a positive column with periodic BCs appear pretty different compared to the striation development between the electrodes previously studied in Argon [21]. In the case of the positive column, striations appear uniformly within the computational domain after ~150 µs, which is the time scale of the order of the ambipolar diffusion time. The waves propagate to the cathode side with velocities of about 100 m/s.

Small amplitude modulation of $n_e(L/2,t)$ oscillations are noticeable in Figure 2 (b). They indicate deviations from resonance conditions for the given value of $U$. Effects of the column length $L$ on the wave properties are briefly discussed in Section IV.

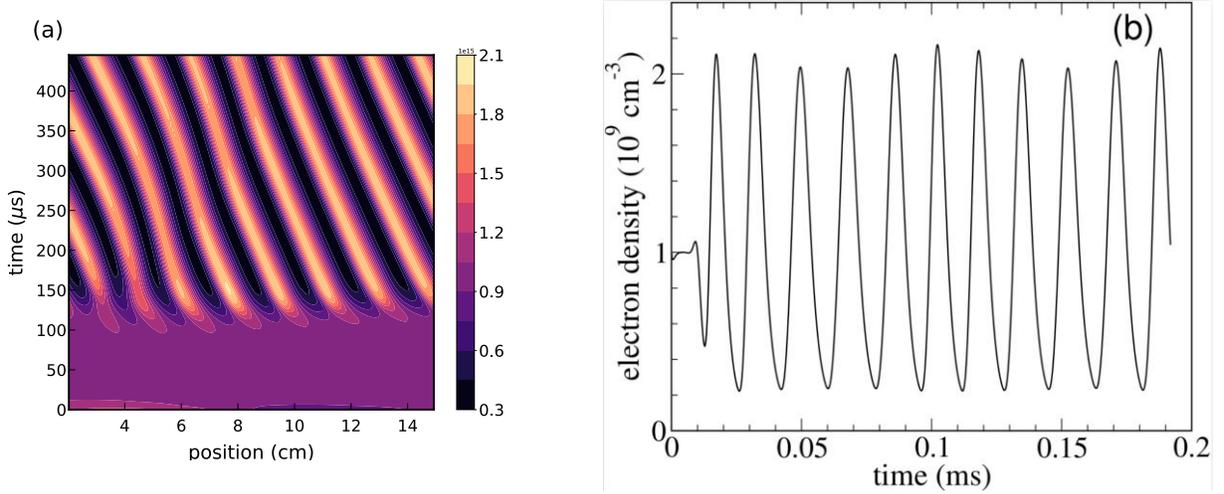

Figure 2: The contours of electron density $n_e(x,t)$ during striation development in the positive column (a) and the corresponding time evolution of electron density in the middle of column (b).

Below, we describe the properties of moving striations in the positive column obtained in our simulations for Neon, Argon, and Helium. We show instantaneous distributions of EEPF and the wave properties established after an initial transient process.

### 1. Three types of waves in Neon

Simulations of Neon plasma were performed for gas pressure $p = 1$ Torr and plasma density $\langle n \rangle = 10^{15}$ m$^{-3}$. By varying the applied voltage $U$, we obtained three types of waves.

#### a. S waves

Figure 3 shows an example of $S$ waves obtained for $L = 3$ cm, $U = 48$ V. These conditions correspond to capillary discharges with $R = 2$ mm and $\langle E \rangle = 16$ V/cm. There are two striations with the potential drop $\Delta\varphi_\Lambda = 24$ V over striation length. The Debye length (~ 0.65 mm) is comparable to the tube radius, and there are notable deviations from quasineutrality. The electron temperature is relatively high, $T_e$ ~6.6 - 8.3 eV (compared to $R$ and $P$ waves described below), and the ionization and excitation rates are of the same order of magnitude. These conditions are close to the boundary of the plasma stratification (1) in Figure 1 (a). The wave frequency is 1 MHz, and the ambipolar particle loss frequency is 3.25 μs$^{-1}$.

Figure 3 illustrates the peculiarities of electron kinetics in phase space. The solid white line shows an instantaneous distribution of the electrostatic potential $\varphi(x,t)$. The two dashed lines indicate Neon's excitation and ionization thresholds ($\varepsilon_1 = 16.62$ and $\varepsilon_i = 21.6$ eV). Since, under our conditions, the electron energy loss in elastic collisions with atoms is negligible, electrons diffuse along surfaces of constant total energy $\varepsilon = u + \varphi(x,t)$ (such as the solid white line in Figure 3). They experience significant energy loss for excitation of atomic level and direct ionization (shown by two solid red lines). Therefore, the length $\lambda_\varepsilon = \varepsilon_1 / (e\langle E\rangle)$ appears as an intrinsic spatial scale for the striation development in noble gases.

Figure 4 shows spatial distributions of macro-parameters calculated from the EEPF. The maximum plasma density appears near the minimum of the electric field strength, slightly shifted to the cathode due to the resistive component of the electric field. The electron temperature reaches a maximum near the maximum of the electric field amplitude. The maximum excitation and ionization rates are between electron density and temperature maxima. The wave moves towards the cathode due to the spatial shift of the ionization rate to the cathode relative to maximum plasma density. The ionization creates new electron-ion pairs producing an "ionization wave" propagating toward the cathode.

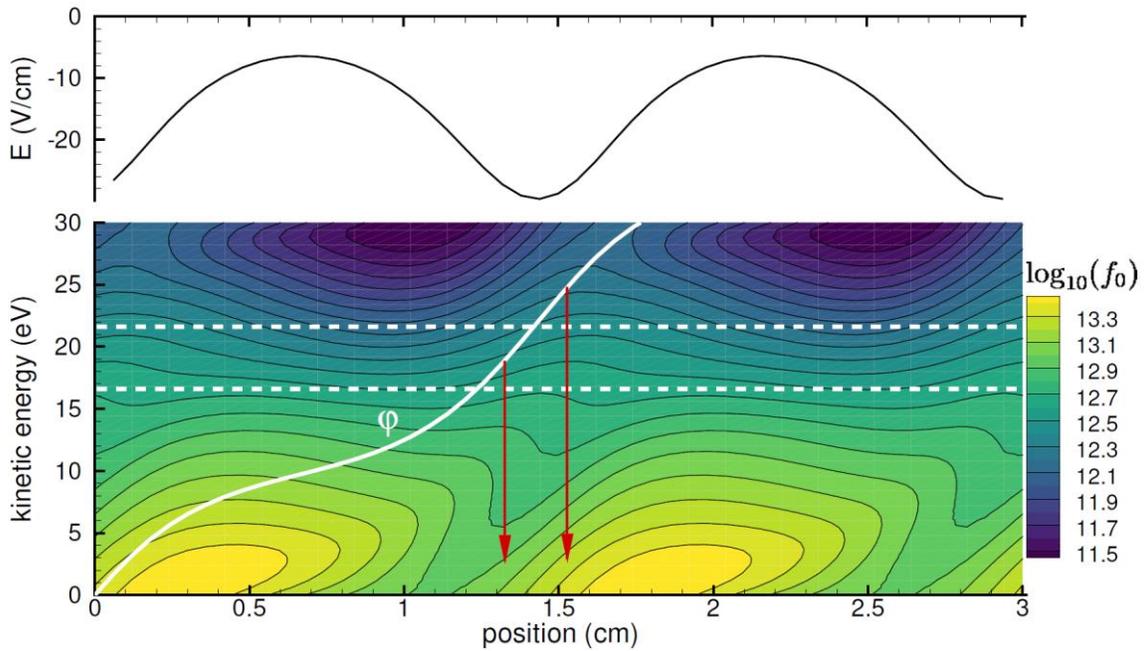

Figure 3: Contours of EEPF and the electric field profile at t = 63 μs. The solid white line shows the electric potential. The dashed lines show the excitation and ionization thresholds.

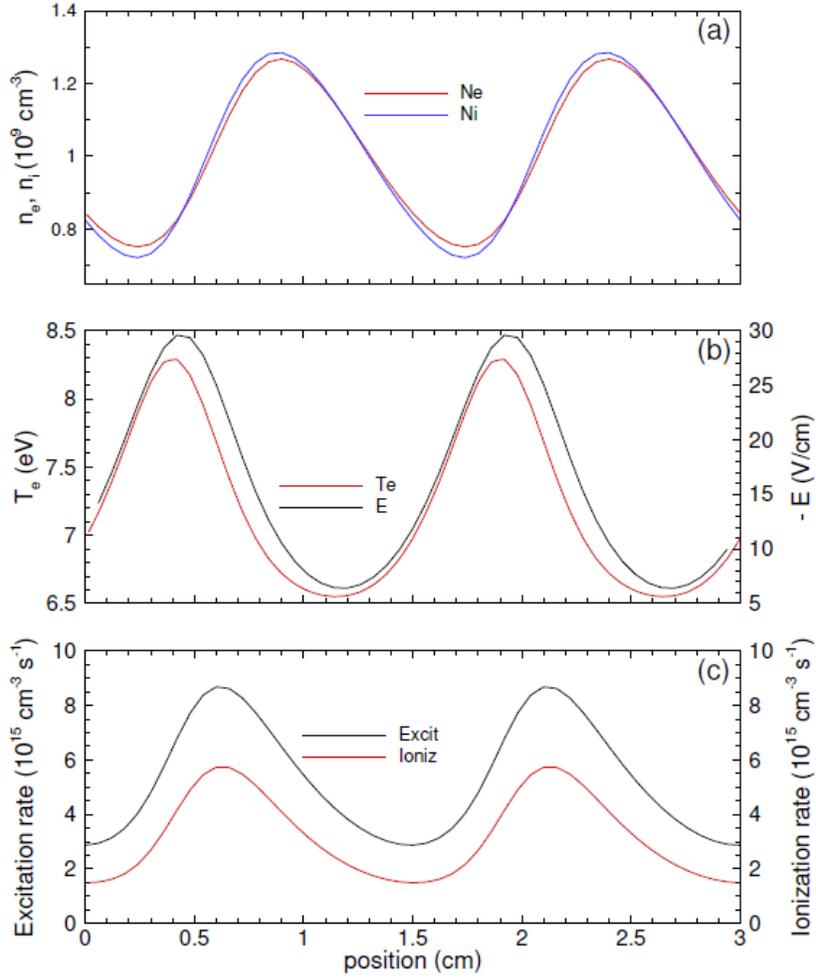

Figure 4: Spatial distributions of electron and ion densities (a), electron temperature and the electric field (b), and excitation and ionization rates (c) for the conditions of Figure 3.

Figure 5 shows an example of nonlinear S waves obtained for $L = 6$ cm, $U = 120$ V. These conditions correspond to $\langle E \rangle = 20$ V/cm and $R \sim 1.6$ mm. The potential drop over striation length is $\varphi_\Lambda = 24$ V again. However, there are electric field reversals, which form groups of electrons trapped in the potential wells. The variations of the electron temperature are more significant (from 5.4 to 9 eV), and the ionization and excitation rates are even closer to each other compared to Figure 3. The EEPF resembles a "local" one, with intense electron heating taking place in the areas of strong electric fields.

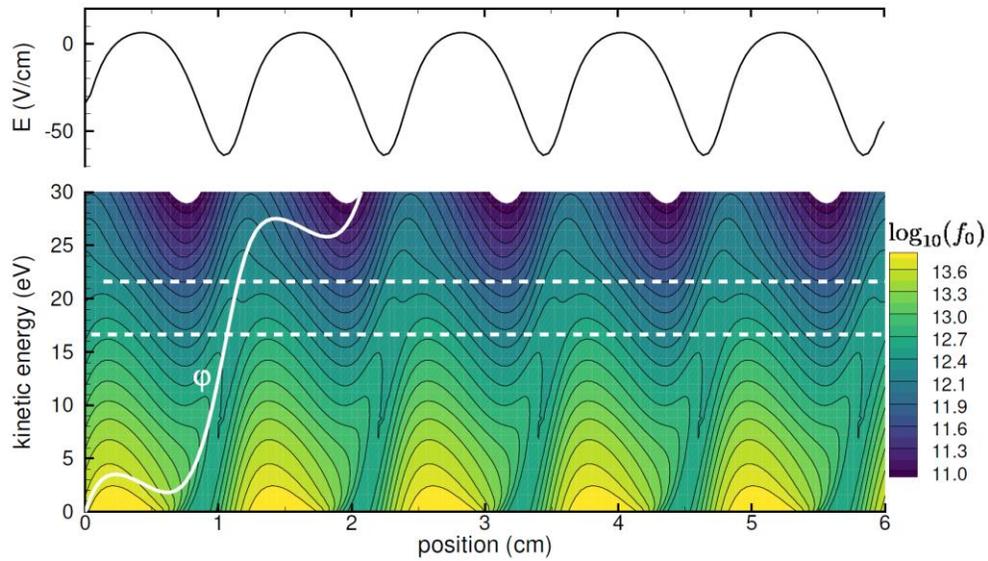

Figure 5: Contours of EEPF and the electric field profile for nonlinear S waves in Neon at t = 93 µs. The solid white line shows the electric potential. The dashed lines show the excitation and ionization thresholds.

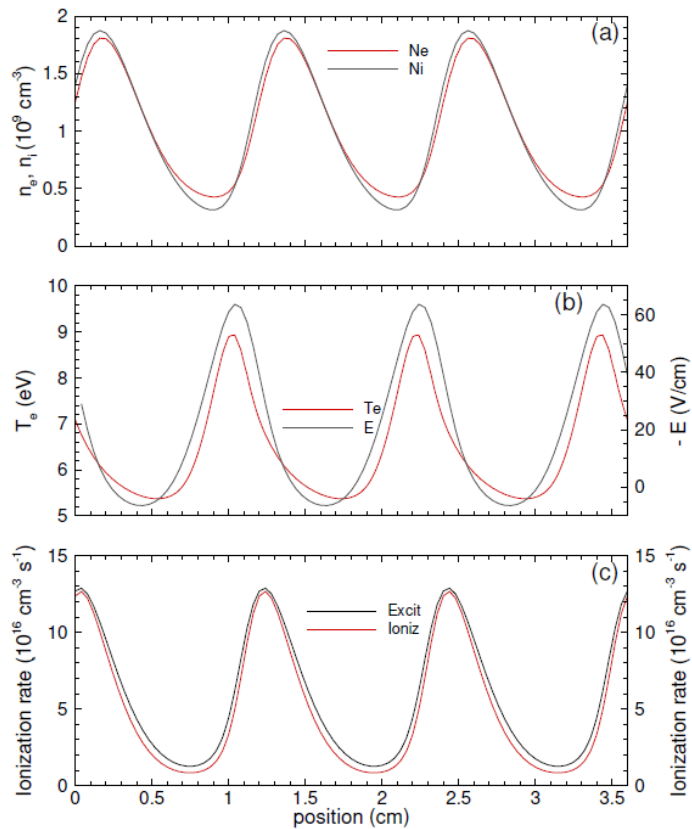

Figure 6: Spatial distributions of electron and ion densities (a), electron temperature and the electric field (b), and excitation and ionization rates (c) for the conditions of Figure 6.

### b. R waves

The R waves were obtained for the electric fields $\langle E \rangle$ in the range 2.53 - 10 V/cm.
Figure 7 shows an example of calculation for $L = 15$ cm, $U = 38$V. These conditions correspond to $R = 1.4$ cm and $\langle E \rangle = 2.53$ V/cm. There are three striations with a wavelength of 5 cm and the potential drop $\Delta\varphi_\Lambda = 12.7$ V over striation length. The Debye length is much smaller than the tube radius, and there are no noticeable deviations from quasineutrality.

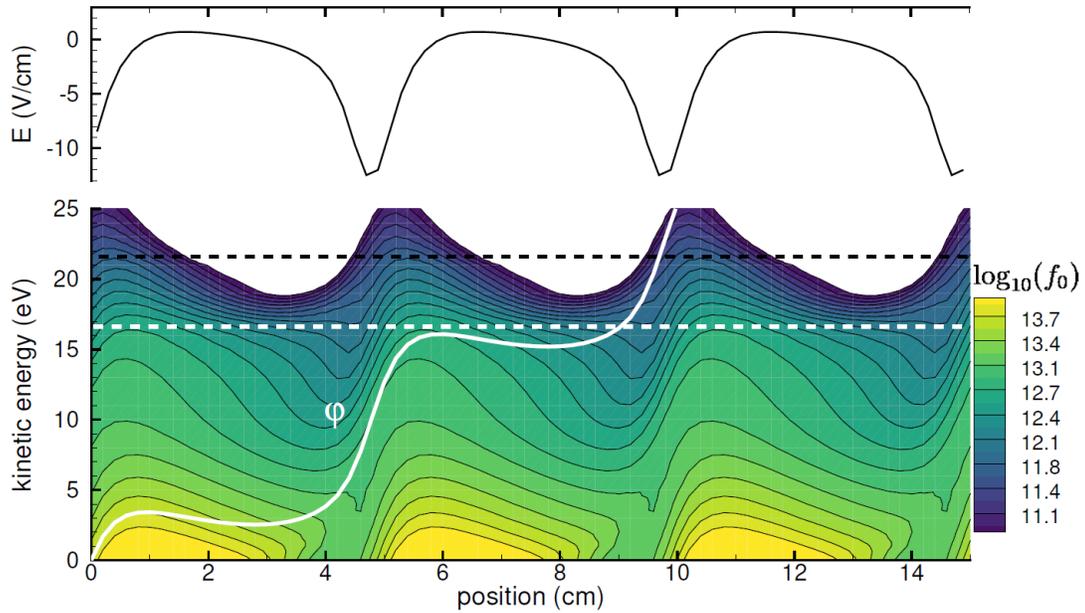

Figure 7: Contours of EEPF and the electric field profile for R waves at $t = 242$ μs. The solid white line shows the electric potential. The dashed lines show the excitation and ionization thresholds.

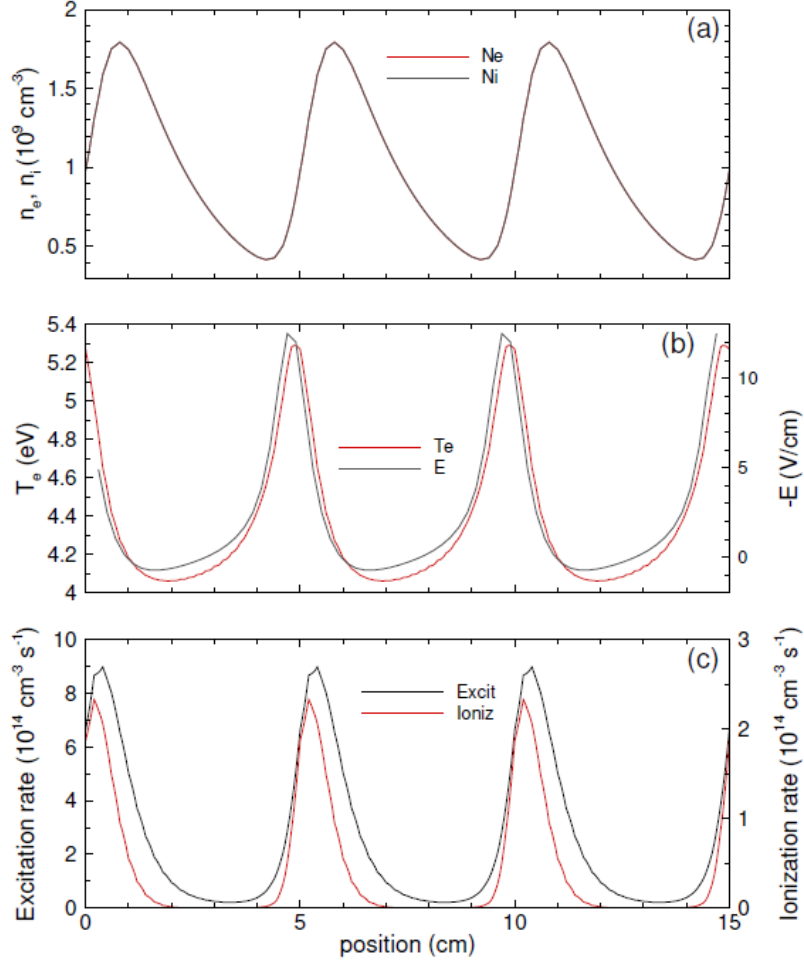

Figure 8: Spatial distributions of electron and ion densities (a), electron temperature and the electric field (b), and excitation and ionization rates (c) for the conditions of Figure 7.

c. P waves

The P waves were obtained for the electric fields in the 1.87 – 2.6 V/cm range. Figure 9 and Figure 10 show an example of P waves for $L = 15$ cm and $U = 28$V, which correspond to $\langle E \rangle = 1.87$ V/cm and $R = 2.2$ cm. The wavelength is 5 cm, and the potential drop over striation length is 9.3 V. The Debye length of ~1.0 mm is much smaller than the tube radius, and there are no noticeable deviations from quasineutrality. The waves are highly nonlinear with strong electric field reversals. The electron temperatures are the lowest, in the range $T_e$ ~ 4.1-5.2 eV, compared to the S and R waves. The excitation and ionization rates are strongly modulated. The maximal value of the ionization rate is lower than the maximal value of the excitation rate by a factor of 5.

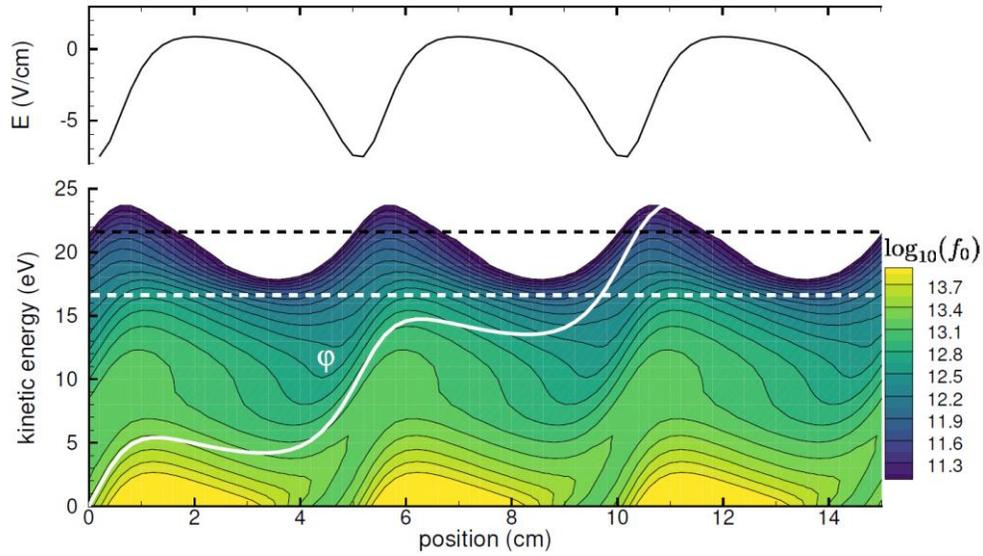

Figure 9: Contours of EEPF and the electric field profile for P waves at t = 1.0 ms. The solid white line shows the electric potential. The dashed lines show the excitation and ionization thresholds.

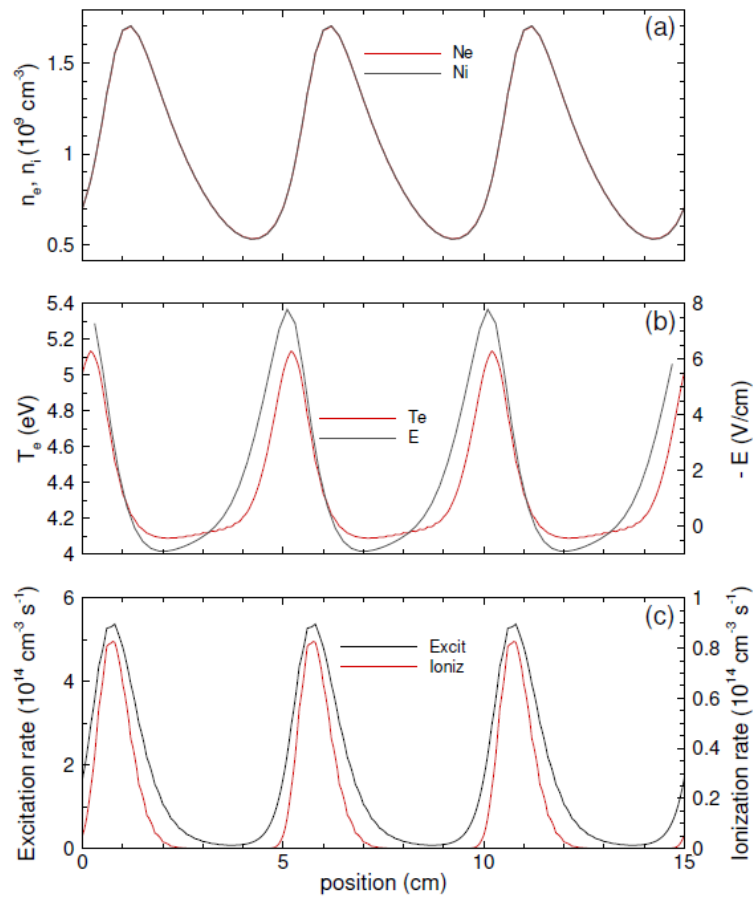

Figure 10: Instantaneous spatial distributions of electron and ion densities (a), electron temperature and the electric field (b), and excitation and ionization rates (c) for the conditions of Figure 9.

Figure 11 summarizes the calculated dependence of the striation length $\Lambda$ on the electric field $\langle E \rangle$, and the dependence of the electric field $\langle E \rangle$ on tube radius $R$. The calculated potential drops for the three types of striations are close to the values observed by Novak [30]. The calculated dependence of the electric field $\langle E \rangle$ on tube radius $R$ is consistent with the experimental observations [7,8].

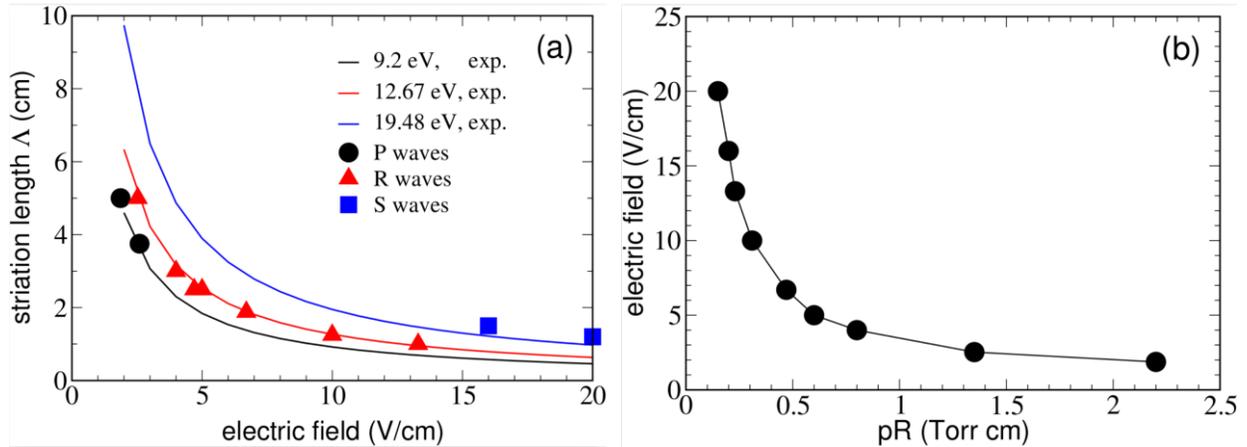

Figure 11: (a) The dependence of striation length $\Lambda$ on the electric field $\langle E \rangle$: Points are our simulations, and solid curves express Novak's law. (b) The calculated dependence of the electric field $\langle E \rangle$ on tube radius $R$ in Neon at 1 Torr.

Similarity laws [12] dictate that the properties of positive column plasma are similar in a given gas for the same values of $pR$, $i/R$, and the wall temperature. Coulomb collisions, stepwise ionization, volume recombination (by pair collisions), and gas heating do not destroy the similarity laws. However, the deviation from quasineutrality and three-particle collisions do. Similarity laws are also valid in the presence of waves in the positive column. According to these laws, the waves with the same reduced frequencies $fR$ (or $f/p$) should have the same reduced length $\Lambda / R$ (or $\Lambda p$). Figure 12 shows the reduced wavelengths $\Lambda / R$ obtained in our simulations. There are three wave types: S waves with $\Lambda/R \sim 7.5$, R waves with $\Lambda/R \sim 4$, and P waves with $\Lambda/R \sim 2.5$. These $\Lambda/R$ values are in good agreement with the experiments [31] where $\Lambda/R \sim 6$ were obtained for S waves, $\Lambda/R \sim 3$ for R waves, and $\Lambda/R \sim 2 \div 3$ for P waves. The experiments were conducted in tubes of radius $R = 1.5$ and 3 cm over a range of discharge currents.

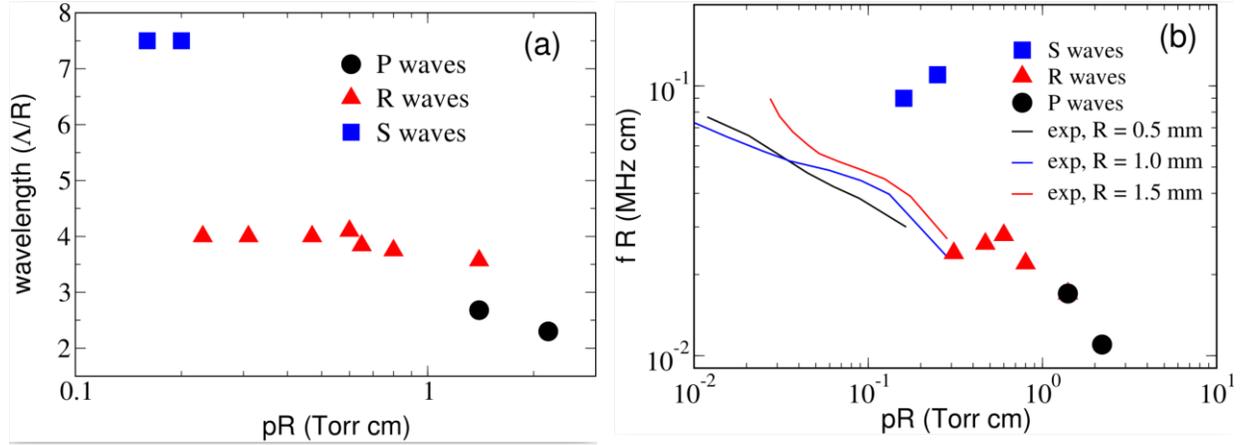

Figure 12: Calculated dependence of the striation length $\Lambda/R$ (a) and frequency $fR$ (b) on $pR$ for S, P, and R waves in Neon at 1 Torr and plasma density $10^{15}$ m$^{-3}$. Solid lines in Figure (b) show the experimental results.

The right part of Figure 12 shows the calculated reduced frequencies $fR$ for the S, P, and R waves and the experimental data obtained in capillary tubes near the boundary of the striation existence at low discharge currents [32]. The violation of the similarity law in the experiments (the dependence of $fR$ on $R$) was attributed [32] to deviations from quasineutrality and to the transition from the ambipolar to free diffusion in the radial direction, which is expected to occur near this boundary. Our simulations (see above) confirm this assumption.

It is seen in Figure 12 that the frequency of the waves increases with decreasing $pR$, as discussed in [32]. The waves in capillary tubes with a radius $R \sim 1$ mm could have frequencies up to 1 MHz. Our simulated frequencies for the R waves agree with the experiment [32]. Only the R waves were observed in these experiments; no S waves could be excited due to experimental limitations. As discussed above, our model is not valid for small $pR$, which corresponds to the conditions when the particle mean free path $\lambda$ exceeds the tube radius $R$. The calculated frequencies for $pR \sim 1$ are about an order of magnitude higher than the experimentally observed in [31]. This discrepancy can be attributed to the neglect of stepwise ionization in our simulations. Our previous work [21] showed that the inclusion of stepwise ionization could decrease the wave propagation velocity by two orders of magnitude due to the transition from ion-guided to metastable-guided waves.

2. **S waves in Argon**

Similar calculations have been performed for Argon at gas pressures 0.25, 0.5, and 1 Torr and plasma density $\langle n \rangle = 10^{15}$ m$^{-3}$ for different electric field values. However, only one type of wave was found in our simulations. Figure 13 and Figure 14 show results for 1 Torr, $L = 5$ cm, $U = 60$ V, which corresponds to $\langle E \rangle = 12$ V/cm. The values of $R = 2.4$ mm, and Debye length $\sim 6.0$ mm were obtained. Highly nonlinear S waves with a length of 1.25 cm and the potential drop $\Delta\varphi_\Lambda = 15$ V were found. The relatively large gap between the excitation and ionization thresholds of

11.55 and 15.8 eV in Argon and the rapid decrease of the EEPF at energies above the excitation threshold (compare Figure 13 with Figures 3, 7, and 9) should make the stepwise ionization more important in Argon compared to Neon.

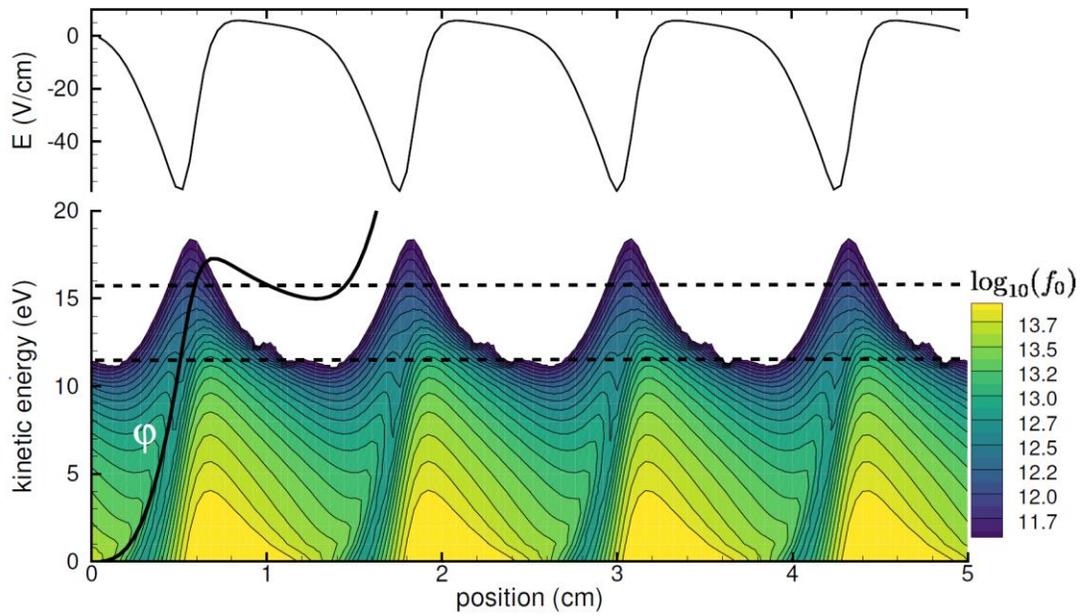

Figure 13: Contours of EEPF and the electric field profile for S waves in Argon at t = 166 µs. The solid black line shows the electric potential. The dashed lines show the excitation and ionization thresholds.

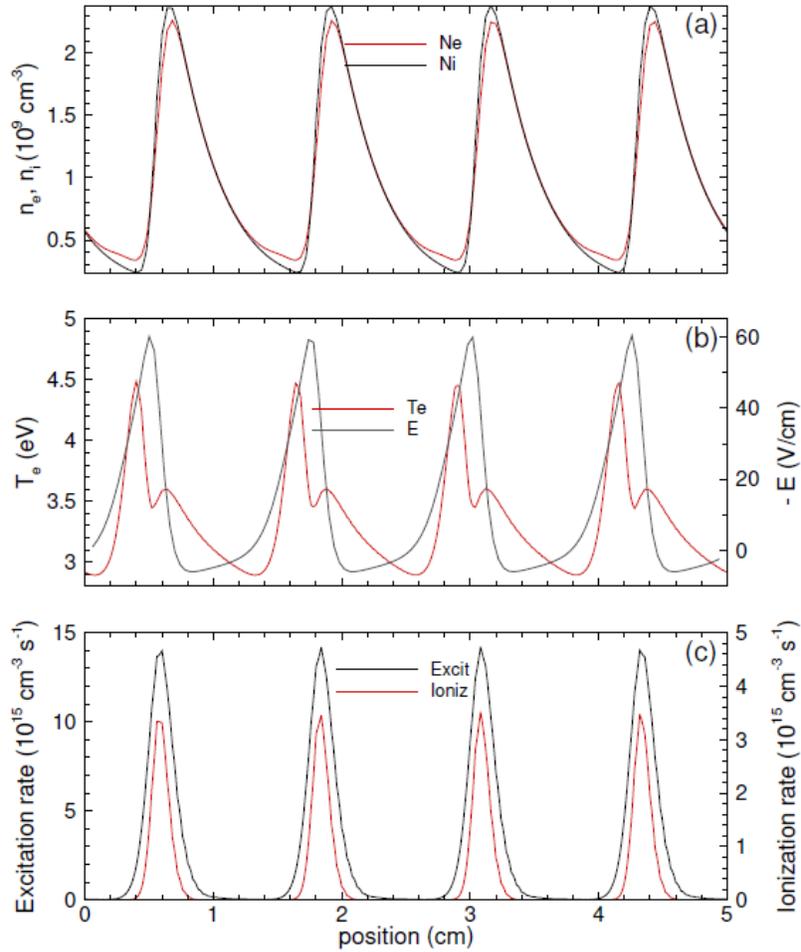

Figure 14: Instantaneous spatial distributions of electron and ion densities (a), electron temperature and the electric field (b), and excitation and ionization rates (c) for the conditions of Figure 13.

Figure 15 summarizes the calculated dependencies of the striation length on the electric field $\langle E \rangle$ and the dependence of $\langle E \rangle$ on the tube radius $R$ for Argon at different pressures. The calculated potential drops for the S waves are close to the values observed by Novak [33]. The calculated dependence of the electric field $\langle E \rangle$ on tube radius $R$ is consistent with the experimental observations [7,8]. We have also observed that the wave frequency increases with increasing $\langle E \rangle$ from 0.1 MHz at 10 V/cm to 1.3 MHz at 40 V/cm at 1 Torr. However, no detailed comparison with experiments has been attempted because it is expected that the P and R waves observed in the Argon appear due to stepwise ionization neglected in our simulations.

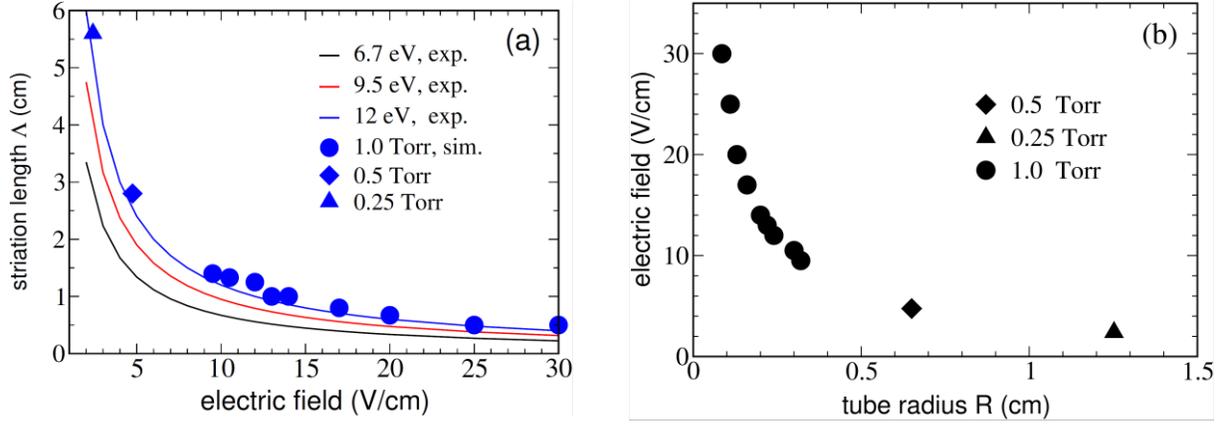

Figure 15: (a) The dependence of striation length on the electric field $\langle E \rangle$, and (b) the dependence of the electric field $\langle E \rangle$ on tube radius $R$ (b) for Argon at different pressures.

### 3. Peculiarities of plasma stratification in Helium

We could not obtain striations in Helium under discharge conditions similar to those in Neon and Argon: the plasma appeared stable against ionization waves. However, by reducing the ion mobility compared to actual values, we could obtain robust striations for up to 1/3 of the real mobility value ($\mu_{i0}$) at at $p = 1$ Torr. Figure 16 shows the instantaneous contours of EEPF, the electric potential, and the electric field profile in a gap $L = 8$ cm and $U = 50$ V. Figure 17 shows the corresponding profiles of plasma density, electron temperature, electric field, excitation, and ionization rates. The calculated tube radius $R = 1.1$ cm, the Debye length is ~ 1 mm, and the potential drop per striation length is $\Delta\varphi_\Lambda = 16.7$ V. No field reversals are observed in this case due to the low amplitude of the waves. Striations become stronger with decreasing ion mobility, and the tube radius $R$ decreases. The speed of the striations increases from 50 m/s to 80 m/s as $\mu_i/\mu_{i0}$ increases from 0.1 to 0.33.

Based on these results, we conclude that the leading cause of the experimentally observed stability of Helium plasma to stratification is due to the lowest mass ratio of ions to electrons, $M/m$. The lowest mass ratio results in high ion mobility of Helium compared to other noble gases. It enhances the rate of the EEPF relaxation in elastic collisions of electrons with atoms and the gas heating by electrons. It should also be noted [7] that peculiarities of the energy dependence of the elastic collision cross-section in Helium result in a near-Maxwellian EEPF in the elastic energy range at higher gas pressures and, therefore, in the absence of the EEPF Maxwellization by Coulomb collisions. All these facts make Helium plasma stratification quite distinct from other gases.

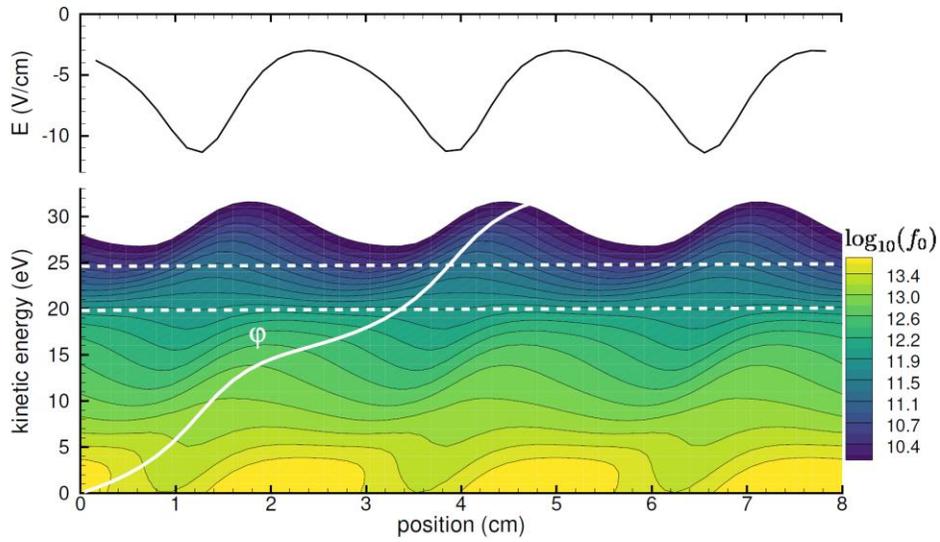

Figure 16: Striations in Helium with ion mobility reduced by 3. Contours of EEPF and the electric field profile at t = 539 μs. The dashed lines show Helium's excitation and ionization thresholds of 19.82 and 24.6 eV.

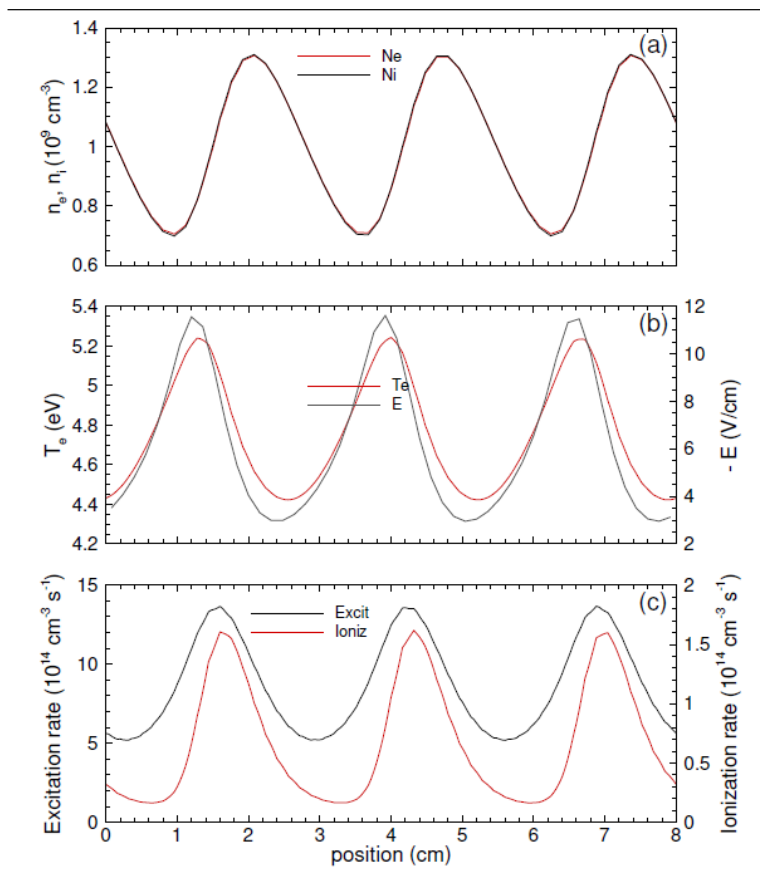

Figure 17: Spatial distributions of electron and ion densities (a), electron temperature and the electric field (b), and excitation and ionization rates (c) for the conditions of Figure 16.

## 4. The Principle of Minimal Power

According to Raizer [34], ionization waves exist because maintaining a striated plasma state is more efficient than a striation-free. The developed model helps substantiate this principle of minimal power for plasma stratification [35]. Figure 18 (a) compares the calculated rates of ionization $I(x)$ and average loss $\langle n(x)\rangle/\tau$ in the striated and striation-free Argon positive column at $p = 1$ Torr for the average plasma density $\langle n \rangle =10^{15}$ m$^{-3}$, and the average electric field $\langle E \rangle = 17$ V/cm. A striation-free column was obtained in our simulations by using one spatial cell, which prevented plasma stratification. In these simulations, the electric field $\langle E \rangle$ was the same in the striated and striation-free plasma. However, the average value of the ionization frequency (which equals the average loss $\langle n(x)\rangle/\tau$ in the striated case is 3.3 times higher than in the striation-free case. Therefore, maintaining the striated plasma is 3.3 times more efficient than the striation-free case. Since the particle loss balances the ionization, the calculated radius of the tube for the striated case ($R = 2.5$ mm) is substantially smaller than the one for the non-striated case ($R = 6.8$ mm).

The calculated ionization enhancement factor in striated plasma is shown in Figure 18 (b) as a function of the electric field $\langle E \rangle$ in Argon and Neon. The ionization enhancement factor is more significant in striated Argon plasma and decreases with increasing $\langle E \rangle$. In Neon, the enhancement factor drops quickly with an increasing electric field. At $\langle E \rangle = 5$ V/cm, there is almost no enhancement, which means that the ionization rate is a linear function of the electric field at these conditions.

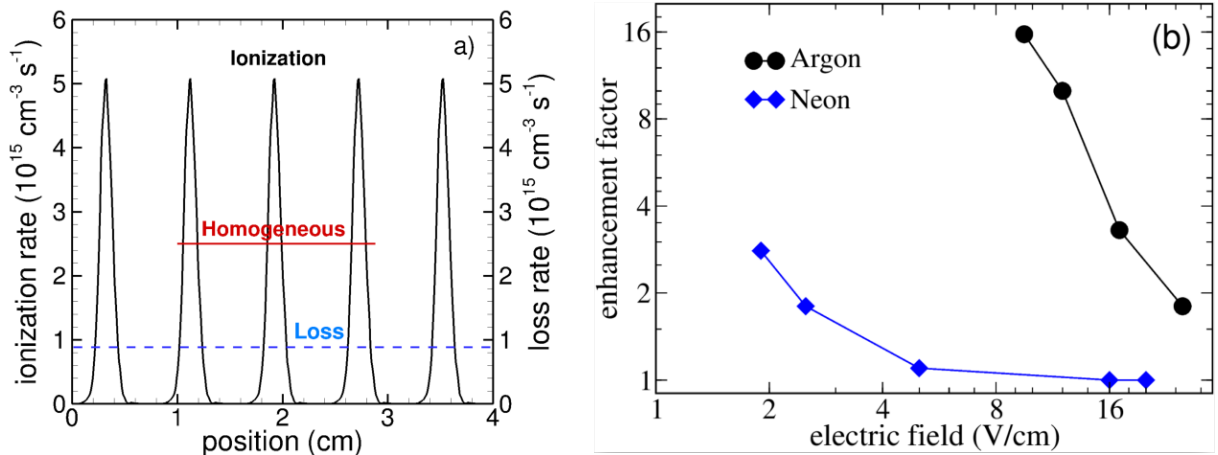

Figure 18: The rates of ionization $I(x)$ (solid black line) and average loss $\langle n \rangle/\tau$ (dashed blue) in stratified and striation-free (red line) Argon plasma (a). The calculated ionization enhancement factors for Argon and Neon as functions of the average electric field $\langle E \rangle$ (b).

Our simulations confirm that the minimal power principle applies to moving striations in DC discharges of noble gases at low currents. Sustaining stratified plasma is more efficient than striation-free plasma when the ionization rate is a nonlinear function of the electric field. However, the nonlinear dependence of the ionization rate on the electric field appears unnecessary for plasma stratification. The case of Neon demonstrates that striations of $S$, $P$, and $R$ types could exist with minimal or no ionization enhancement.

## IV. Discussion

Our simulations confirmed that plasma stratification in DC discharges in noble gases at low $pR$ and $i/R$ is due to nonlocal electron kinetics and the intrinsic spatial scale $\lambda_\varepsilon = \varepsilon_1/(e\langle E \rangle)$ over which the electrons gain kinetic energy equal to the excitation threshold of atoms $\varepsilon_1$. When the spatial length of the EEPF relaxation is considerable compared to $\lambda_\varepsilon$, electrons diffuse in phase space with conservation of their total energy, and the spatial gradients in the FP kinetic equation for electrons dominate. The EEPF relaxation in noble gases occurs via the energy loss of electrons in elastic collisions with atoms, excitation of several atomic levels, and generation of secondary electrons in the ionization events [10]. Under conditions studied in the present paper, the energy loss in inelastic collisions dominates, and the generation of secondary electrons in the ionization events is the primary channel of the EEPF relaxation.

### 1. Effect of column length on wave properties

To illustrate the effect of the positive column length $L$ on the wave properties, we conducted simulations for $L = 6$ cm, $U = 96$ V, $\langle n \rangle = 10^{15}$ m$^{-3}$ that correspond to the same value of $\langle E \rangle = 16$ V/cm as in Figure 3. Figure 19 and Figure 20 show nonlinear S waves with electric field reversals obtained for the same radius $R$ of 2 mm and the Debye length of 0.7 mm, as in Figure 3. However, the nonlinear waves are shorter, and the potential drop $\Delta\varphi_\Lambda = 19.2$ V is lower compared to the linear waves shown in Figure 3. The wave frequency of the nonlinear waves (0.5 MHz) is half of the frequency of the linear waves in Figure 3.

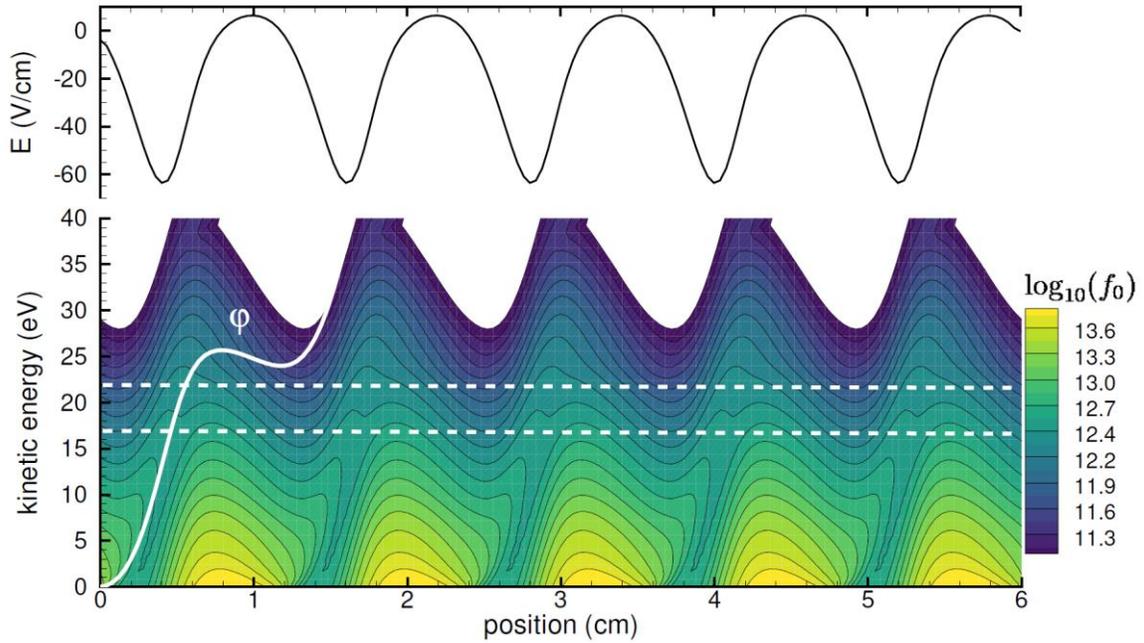

Figure 19: Contours of EEPF and the electric field profile for nonlinear $S$ waves corresponding to the same value of $\langle E \rangle = 16$ V/cm as in Figure 3.

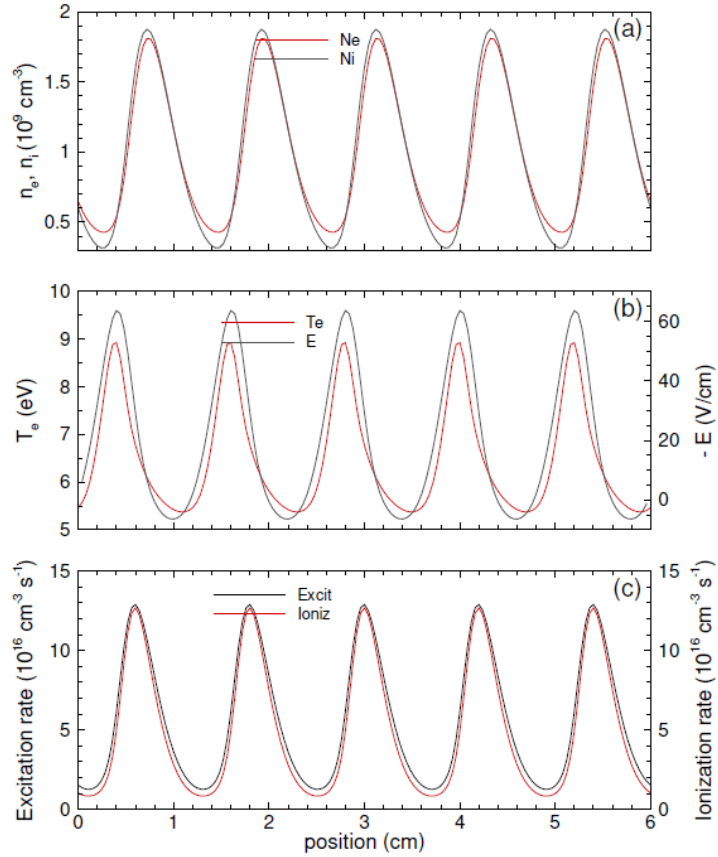

Figure 20: Spatial distributions of electron and ion densities (a), electron temperature and the electric field (b), and excitation and ionization rates (c) for nonlinear $S$ waves shown in Figure 19.

## 2. Effects of plasma density

To illustrate the effects of plasma density on the wave properties, we performed calculations for the same values of $L = 3$ cm and $U = 48$ V (as in Figure 3) and $L = 6$ cm, $U = 96$ as in Figure 20, but with an order of magnitude lower plasma density $\langle n \rangle = 10^{14}$ m$^{-3}$. No striations formed for the first case ($L = 3$ cm, $U = 48$ V). However, robust striations appeared in the second case ($L = 6$ cm, $U = 96$ V). The calculated tube radius $R = 1.8$ mm is smaller than the Debye length ~2 mm, with pronounced deviations from quasineutrality (Figure 21).

Figure 21 shows the EEPF and the electric field profile. The electric field resembles the spatially modulated field used in the previous publications [9, 36], with a modulation amplitude of about 100% in our case. The EEPF corresponds to S waves with the potential drop $\Delta\varphi_\Lambda = 24$ V, similar to the linear waves in Figure 3. However, the wave frequency (0.3 MHz) is three times lower than in the case of higher plasma density in Figure 3.

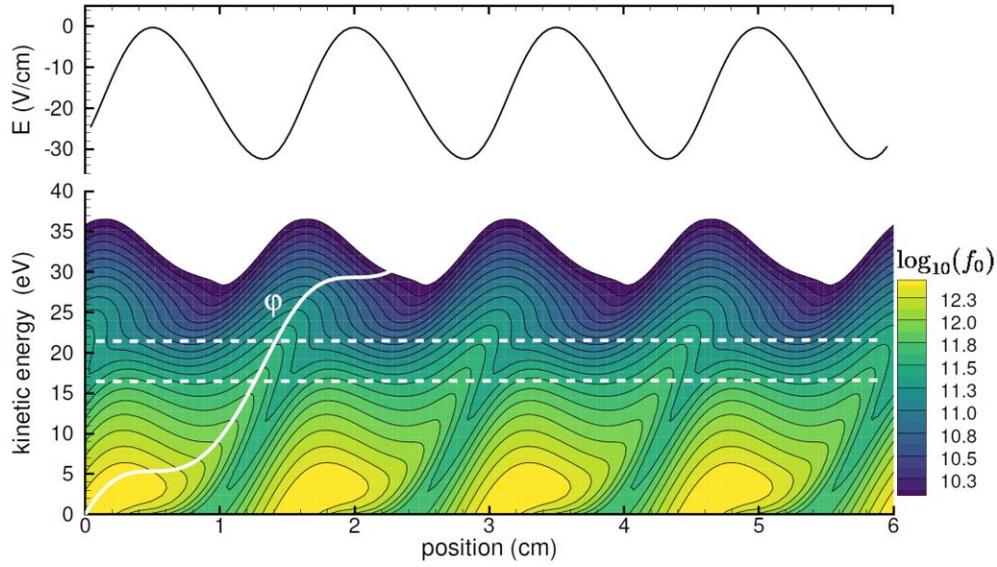

Figure 21: Contours of EEPF and the electric field profile for linear S waves in Neon at the average plasma density $\langle n \rangle = 10^{14}$ m$^{-3}$.

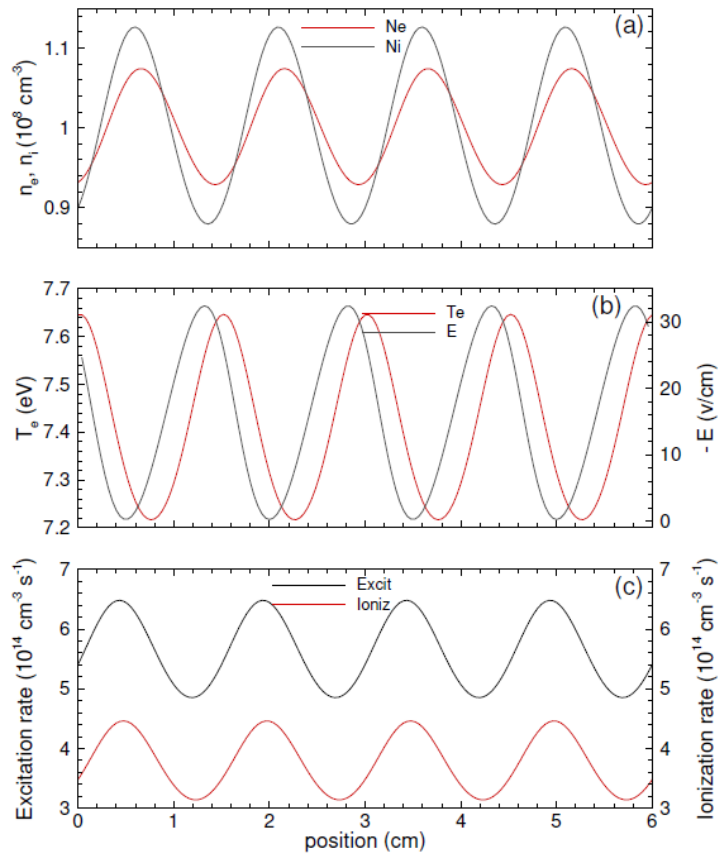

Figure 22: Spatial distributions of electron and ion densities (a), electron temperature and the electric field (b), and excitation and ionization rates (c) for the conditions of Figure 21.

The EEPFs shown in Figure 3 and Figure 21 have the typical characteristic of nonlocal behavior under conditions when the electron energy relaxation length exceeds the stratum length [9]. There is a noticeable shift in coordinates and pronounced peaks propagating along the lines of constant total energy in phase space. Figure 23 illustrates the EEPF dynamics at a given point showing these peaks quantitatively.

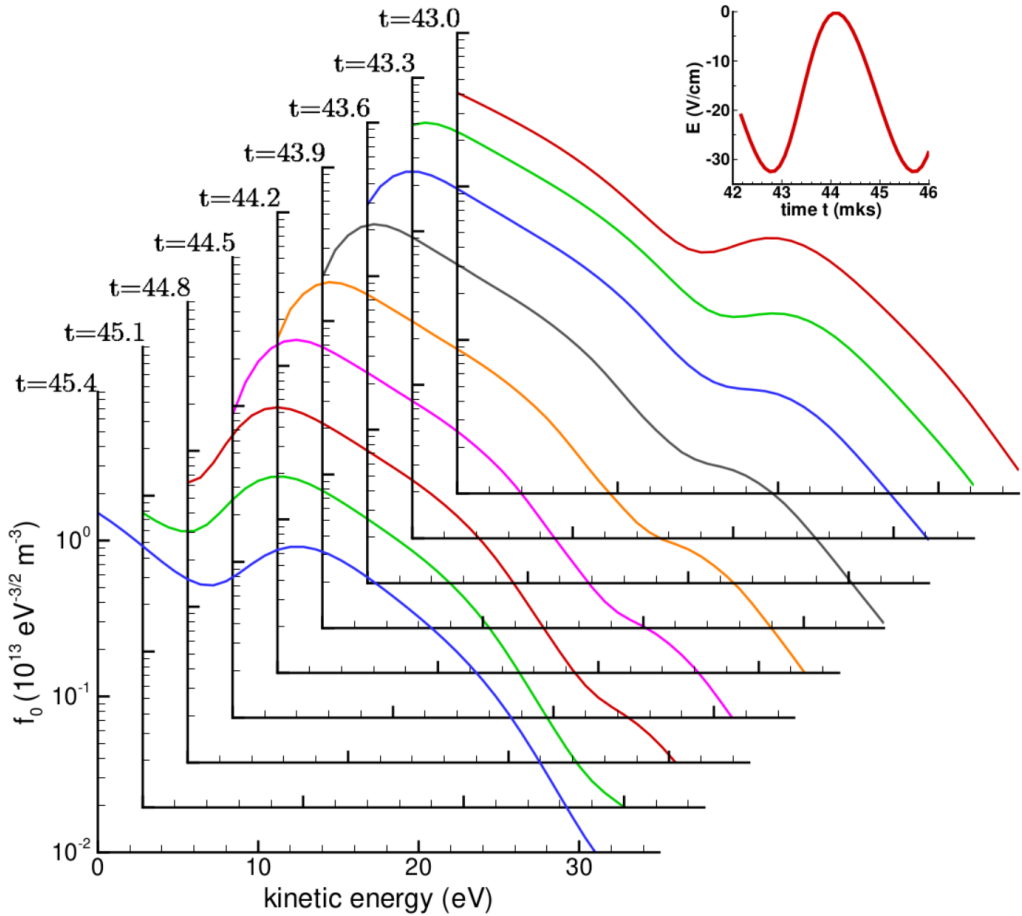

Figure 23: The EEPF dynamics at $x = L$ during one period of the S waves.

### 3. Peculiarities and limitations of the model

Our one-dimensional plasma model has a simplified treatment for radial electron loss. In the actual 2D discharges, only electrons with kinetic energy exceeding the local value of the radial trapping potential can escape to the wall. However, the simplified model for radial electron transport is sufficient for analyzing plasma stratification under the discharge conditions of the present paper. The reason is that the (slow) ambipolar diffusion controls the radial electron loss. Therefore, only small electron multiplication is needed to compensate for the radial loss of electrons. No surprise, we found that variation of the electron loss term had little effect on the striation properties.

The frequency of ionization waves in DC discharges is low compared to the characteristic time scale of the EEPF relaxation. As a result, the EEPF is quasi-static, and the electron temperature has maximal values in the areas of large electric fields. At the same time, the velocity of the waves is large compared to the ion drift velocity. Therefore, the ion motion in the axial direction has little effect on wave propagation. However, ion mobility controls the rate of ambipolar loss of charged particles to the wall, which explains the impact of the ion mobility of plasma stratification in Helium.

Our model neglects stepwise ionization and Coulomb interactions among electrons. These are the two most important processes to be included in the model before a detailed comparison with experiments. Stepwise ionization is responsible for metastable-guided and ion-guided ionization waves observed experimentally. Adding stepwise ionization will extend the applicability of model to higher values of *pR*. The Coulomb collisions are necessary for adequately calculating low-energy electrons, especially when electric field reversals are present. Adding Coulomb collisions is required before a detailed comparison with the experimentally measured EEPFs can be performed. They will also extend the model to higher values of *i/R*.

The FP kinetic equation (1) for EEPF $f_0(r, u, t)$ is valid when the electron mean free path $\lambda$ exceeds the tube radius *R*. When $\lambda$ becomes comparable to the tube radius *R*, the solution of the complete Boltzmann equation for $f_e(r, v, t)$ is required. Simulations of striations experimentally observed at low values of *pR* (below the dashed lines $\lambda = R$ in Fig. 1) require such an extension [2].

## V.  Conclusions

We have obtained moving striations in DC discharges in noble gases (Ne, Ar, He) using a self-consistent hybrid model of collisional discharge plasma at low currents. A 2D kinetic equation for EEPF in the (*x,u*) phase space was solved self-consistently, together with the drift-diffusion equation for ions and the Poisson equation for the electric field. Using the grid-based, non-statistical method for solving the FP kinetic equation for electrons, we could detect striations at much lower values of *E/p* compared to the PIC simulations [22]. Large-amplitude striations have been obtained for *E/p* as low as 9.5 V/cm/Torr in Argon. Our striation onset values for *E/p* are thus in closer agreement with the linear theory [22].

The 2-level excitation-ionization model has been applied at low plasma densities when the nonlinear effects due to gas heating, stepwise ionization, and Coulomb interactions among electrons can be neglected. Moving striations of the *S*, *P*, and *R* types were obtained in Neon for average plasma density $10^{15}$ m$^{-3}$, and a range of the *pR* values corresponding to experiments. Only *S* waves were found in Argon for similar discharge conditions. No striations appeared in Helium for the realistic value of the ion mobility. However, *S* striations have been obtained in Helium by artificially decreasing the ion mobility by a factor of three. Therefore, the reasons for the experimentally observed stability of Helium plasma to ionization waves have been clarified.

We have shown that in Argon, lower average electric fields can maintain stratified plasma compared to striation-free plasma. This result substantiates the principle of minimal power for the kinetic striations in DC discharges in Argon. However, the nonlinear dependence of the ionization rate on the electric field appears unnecessary for plasma stratification. In our simulations, striations of *S*, *P*, and *R* types in Neon appeared with minimal or no ionization enhancement.

We have studied the stratification of a finite-length positive column using periodic boundary conditions similar to [22]. The previous work [28] considered the positive column as a cavity resonator containing a set of modes representing strata of various types. The S-, P-, and R-striations were interpreted as resonances based on the analysis of electron kinetics in a spatially modulated electric field. In our studies, we solved the problem self-consistently, including electron kinetics, ions transport, and the electric field for a range of *pR* values covering all three striation types. Effects of column length on the wave properties have been demonstrated in our simulations. However, this subject deserves further studies to connect the cavity resonator effects with electron bunching [9] and kinetic resonances [36] in spatially modulated electric fields. We expect that exciting results can be obtained for kinetic resonances (when the length of the stratum and the resonant length coincide) and hysteresis phenomena associated with changing the integer number of waves inside the column.

Further development of the model will include metastable atoms and stepwise ionization to explain the two types of experimentally observed ion-guided (fast) and metastable-guided (slow) ionization waves. We also plan to add Coulomb collisions to study the transition between the wave types with increasing the plasma density (discharge current).

## Acknowledgments

This work was supported by NSF projects OIA 1655280 and OIA-2148653, and DOE project DE-SC0021391. VIK thanks Prof. Golubovskii for the valuable discussions.